\documentclass[apl,twocolumn, amsmath,amssymb]{revtex4}
\usepackage{graphicx}
\usepackage{dcolumn}
\usepackage{bm}
\usepackage{color}
\newcommand{\ba}{\begin{eqnarray}}
\newcommand{\ea}{\end{eqnarray}}

\begin{document}
\title{Spin Lifetime in Small Electron Spin Ensembles Measured by Magnetic Resonance Force Microscopy}
\author{K.C. Fong}
\affiliation{Applied Physics, California Institute of Technology, MC 128-95, Pasadena, California 91125}
\affiliation{Department of Physics, Ohio State University, 191 West Woodruff  Ave., Columbus OH 43210}
\author{M.R. Herman}
\author{P. Banerjee}
\author{D.V. Pelekhov}
\author{P.C. Hammel}
\affiliation{Department of Physics, Ohio State University, 191 West Woodruff  Ave., Columbus OH 43210}
\date{\today}
\begin{abstract}
Magnetic Resonance Force Microscopy can enable nanoscale imaging of spin lifetime.  We report temperature dependent measurements of the spin correlation time $\tau_m$ of the statistical fluctuations of the spin polarization---the spin noise---of ensembles containing $\sim 100$ electron spins by this technique. Magneto-mechanical relaxation due to spin-cantilever coupling was controlled and spurious mechanisms that can affect the spin correlation time of the microscopic signal were characterized. These measurements have ramifications for optimizing spin sensitivity, understanding local spin dynamics and for nanoscale imaging.
\end{abstract}
\maketitle

Magnetic Resonance Force Microscopy (MRFM) \cite{s:rmp, h:MagHandbook} can detect magnetic resonance from very small spin ensembles with single electron spin sensitivity \cite{r:singlespin}. For small spin ensembles, statistical fluctuations of the net spin polarization \(P_{\rm net} = (N_{\uparrow} - N_{\downarrow}) / (N_{\uparrow} + N_{\downarrow})\) \cite{r:maminStatisticalPolarization, r:NoiseNMRprl2007} exceed the Boltzmann polarization. Spin noise is a topic of intrinsic interest \cite{Crooker:spinnoise.nature04} as it reveals fundamental information about the microscopic environment around the measured spins. Spin relaxation provides a  powerful approach to probing electronic, magnetic and structural dynamics in materials \cite{slichter}, and plays an important role in Magnetic Resonance Imaging (MRI) where $T_1$- and $T_2$-weighting are used to enhance image contrast \cite{Haacke:MRI}. The signal-to-noise ratio (SNR)---singularly important for high resolution MRFM imaging---is centrally influenced by spin lifetime because it determines the detection bandwidth.

Here, we report measurements of $\tau_m$ in nanoscale ensembles containing $\sim $100 electron spins.  The number of resonant spins and the correlation time $\tau_m$ of their fluctuations are characterized in MRFM experiments by the spectral weight and linewidth, respectively. Ideally, the spin-lattice relaxation time in the rotating frame, $T_{1\rho}$, determines $\tau_m$ \cite{mozyrsky:CantRxn, r:CantileverInducedNuclearRelaxation}. We present systematic measurements of the evolution of  $\tau_m$ with spin modulation depth,  microwave power and sample temperature. We argue, based on these data, that the relaxation time we measure in these experiments are due to intrinsic mechanisms.

Care must be taken to avoid artificially shortening the spin correlation time through mechanisms of technical origin such as the higher order cantilever oscillation modes \cite{r:hundredspin, mozyrsky:CantRxn, r:CantileverInducedNuclearRelaxation}, violation of adiabaticity \cite{r:wago98, r:MaminStatisticalNMR}, and low frequency fluctuations of the field of the micromagnetic probe \cite{r:magflucttheory}.  We avoided these by using mass-loaded cantilevers \cite{r:MassLoadedCantilever}, large cantilever oscillation magnitudes $x_{\rm pk}$ and large transverse oscillating magnetic fields $H_1$. We find that the temperature dependence of $1/\tau_m (T)$ are intrinsic to the sample and are well explained by phonon mediated relaxation processes.

\begin{figure}
\includegraphics[width=\columnwidth]{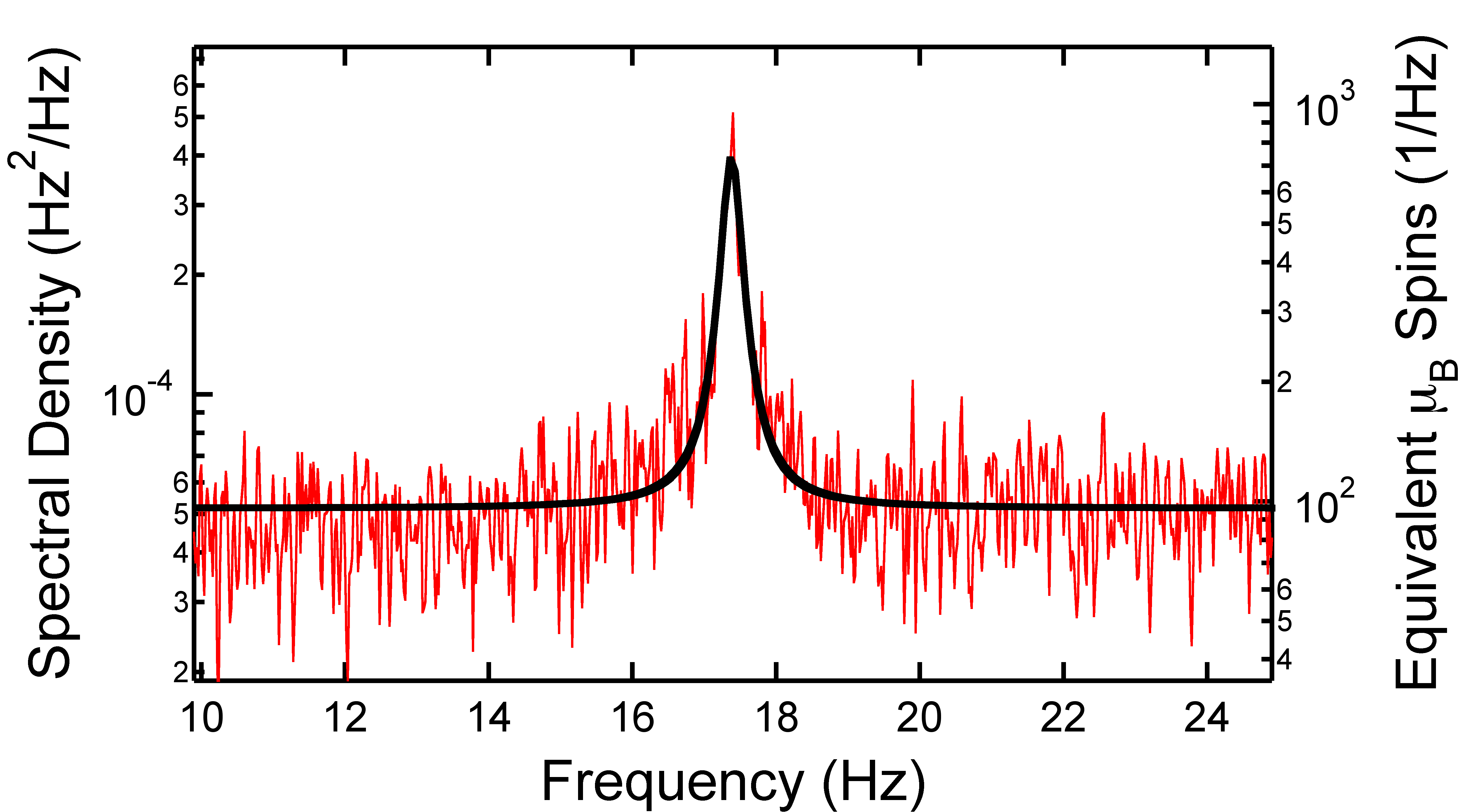}
\caption{Power spectral density of the spin noise. The full width at half maximum and the area of the fitted Lorentzian are 0.30 Hz and 162 mHz$^2$ respectively. They correspond to $\tau_m = 1.06$ s and signal energy of 534 aN$^2$, which is equivalent to 302 electron spins in 80 nm$^3$ voxel with 1.3 G/nm field gradient. Data taken with tip-sample separation $d$ = 350 nm.}
\label{fig:SpinNoise}
\end{figure}
Our experiments were performed in vacuum between 4.2 and 40 K on an optically polished piece of vitreous silica  (see Ref.~\onlinecite{Fong:dissertation}  for details). We measure electron spins present at a density of $\sim 6 \times 10^{17}$ cm$^{-3}$. These spins reside in silicon dangling bonds associated with oxygen vacancy defects known as E' centers, which are produced by $^{60}$Co gamma irradiation  \cite{Castle:EprimePhysRev1963, Castle:silica.jap1965, Castle:PhysRev1965, Yip:OxygenVacancyModel, Weil:ESRinQuartzReview}. The sample is thermally anchored to a temperature controlled copper block. The IBM-style ultrasoft cantilever we used has a spring constant $k \simeq 0.1$ mN/m and a mass-loaded tip to suppress tip motion \cite{r:MassLoadedCantilever} arising from thermal excitation of higher order cantilever oscillation modes \cite{mozyrsky:CantRxn}. The probe magnet is a SmCo$_5$ particle glued to the cantilever and ion-milled to a tapered end whose size is about $300 \times 600$ nm$^2$. It has coercivity and anisotropy fields greater than 1 T at low temperature, thus avoiding spin relaxation induced by fluctuations of the probe magnetic field \cite{r:magflucttheory}. The cantilever frequency $f_c$ is 3062.15 Hz with the tip attached. The transverse oscillating (2.162 GHz) magnetic field $H_1$ is generated by a superconducting microwave resonator \cite{r:scresonator}. The experiments were performed with no external magnetic field applied.

We used the iOSCAR protocol \cite{r:maminStatisticalPolarization} to excite magnetic resonance and measured the resulting cantilever frequency shift $\delta f_c$ resulting from the modulated magnetic interaction between the electron spins and the micromagnetic probe on the cantilever. Random and uncorrelated spin noise leads to a Lorentzian frequency dependence of the power spectral density $S_{f_c}$ of these frequency shifts (see Fig.~\ref{fig:SpinNoise}) as in the random telegraph signal model \cite{r:singlespin, Davenport:RandomSignal}:
\begin{equation} \label{eqn:SpinNoiseLorentzianDSB}
  S_{f_c} = \frac{2\tau_m\epsilon_f}{1 + 4\pi^2\tau_m ^2 (f - f_{m})^2}
\end{equation}
where $f_{m}$ is the iOSCAR modulation frequency and $\epsilon_f$ is the average frequency-signal energy from $N$ resonant electron spins.  The area under the Lorentzian in Fig.~\ref{fig:SpinNoise} is 162 mHz$^2$; this gives a force signal energy $\epsilon$ of 534 aN$^2$; the two are related by
\(\epsilon = (\pi k x_{\rm pk}/2 f_c)^2 \epsilon_f\) \cite{Berman:OSCAR, r:maminStatisticalPolarization}. The measured tip field gradient is about 1.3 G/nm, so the statistical polarization is due to approximately 302 electron spins ($\sqrt{N}$ = 17.4) in a $\sim (80 \;\rm nm)^3$ detected volume. The noise floor, 13 aN/$\sqrt{{\rm Hz}}$, is primarily due to thermal force noise and corresponds to spin sensitivity of $\sim 100$ electrons in a 1 Hz bandwidth. Hereafter both $\tau_m$ and $\epsilon$ are taken from a fit to the single side-band power spectral density obtained by means of a software lock-in amplifier with a bank of low pass filters \cite{r:singlespin, r:NoiseNMRprl2007} to improve SNR. Most of the data points take about 1 hour for averaging.

\begin{figure}
\includegraphics[width=\columnwidth]{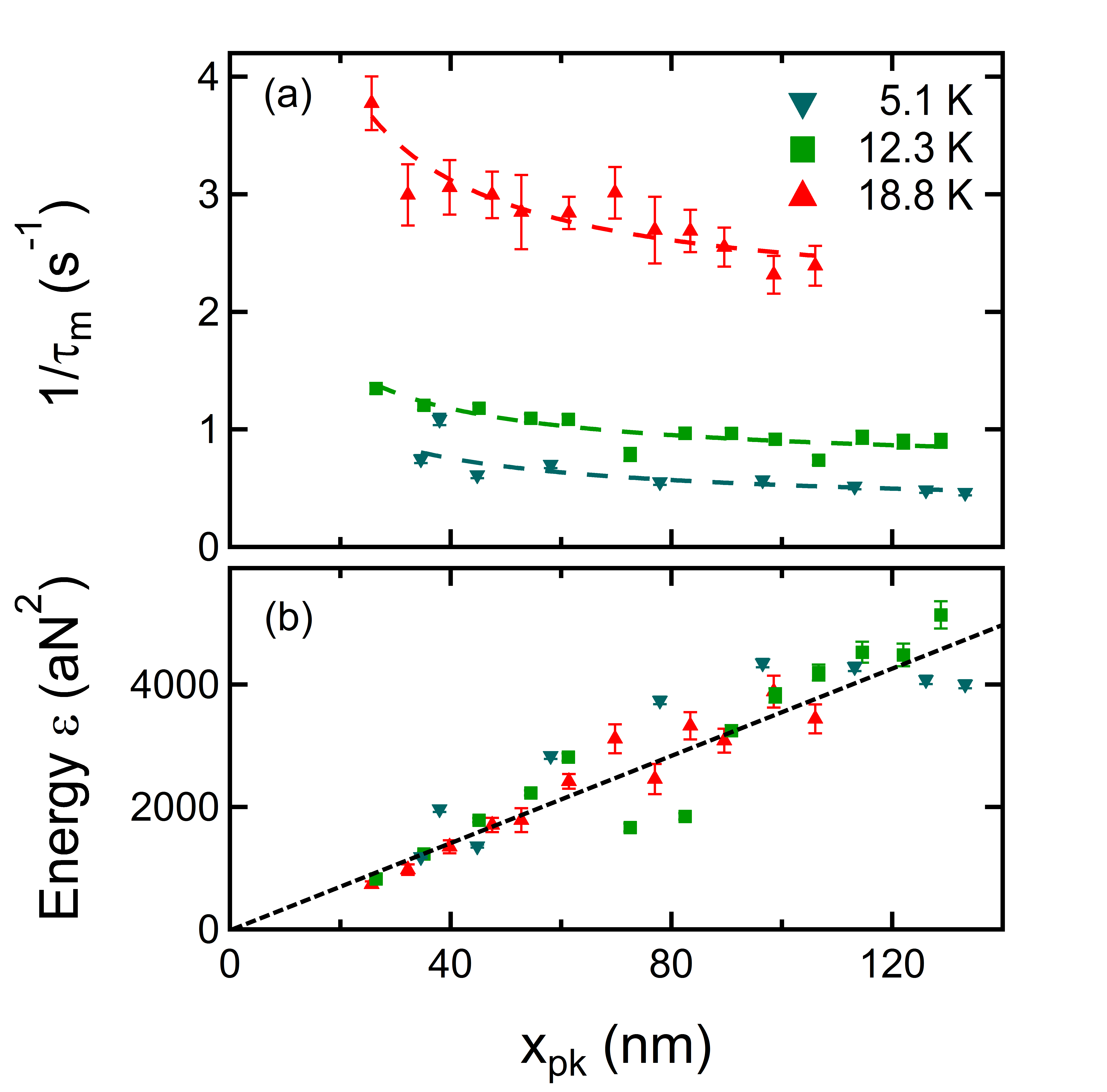}
\caption{Dependence of relaxation rate $1/\tau_m$ on cantilever oscillation amplitude $x_{\rm pk}$ at three different temperatures. $1/\tau_m$ decreases asymptotically toward $T_{1}^{-1}(T)$  as $x_{\rm pk}$ increases. The dashed lines are the fits to phenomenpower law behavior $1/\tau_m = \beta(T) x_{\rm pk}^\alpha + T_{1}^{-1}(T)$ where $-1 < \alpha < -0.7$. Lower panel: the signal energy increases linearly with $x_{\rm pk}$ as expected. In this experiment, $d$ = 200 nm, $P_{\mu w}$ = 2.51 mW and $f_{m}$ = 45 Hz for T = 18.8 K, 19.9 Hz otherwise.}
\label{fig:RelaxationNXpk}
\end{figure}
The correlation time $\tau_m$ is determined by the relaxation time in the rotating frame $T_{1\rho}$ averaged over the distribution of effective field frequencies $\omega_{\rm eff}$ experienced during an adiabatic inversion cycle  \cite{mozyrsky:CantRxn, r:CantileverInducedNuclearRelaxation}. In the absence of excess low frequency spin fluctuations, $T_{1\rho}$ approaches $T_1$ \cite{Ailion:UltraslowMotions, JacquinoGoldman:RotatingFrameRelaxation}. If the spin spends most of its time far off-resonance, that is, if either the microwave frequency modulation or the product of the cantilever oscillation amplitude and the probe gradient is large (ensuring that the extremum of the time-varying effective magnetic field in the rotating frame $H_{\rm eff}$ is much larger than $H_1$), and if the adiabatic condition is satisfied, then $\tau_m$ in iOSCAR should approach $T_1$.

We explored the dependence of  $1/\tau_m$ on $x_{\rm pk}$ at three temperatures (see Fig.~\ref{fig:RelaxationNXpk}). Similar to \cite{JacquinoGoldman:RotatingFrameRelaxation}, we find $1/\tau_m$ decreases asymptotically to the temperature dependent intrinsic relaxation rate: $1/\tau_m = \beta(T) x_{\rm pk}^\alpha +T_{1}^{-1}(T)$  where $-1 < \alpha <-0.7$ (dashed lines). As the resonant slice sweeps through larger volumes with increasing $x_{\rm pk}$, $\epsilon$ increases linearly (lower panel).

\begin{figure}
\includegraphics[width=\columnwidth]{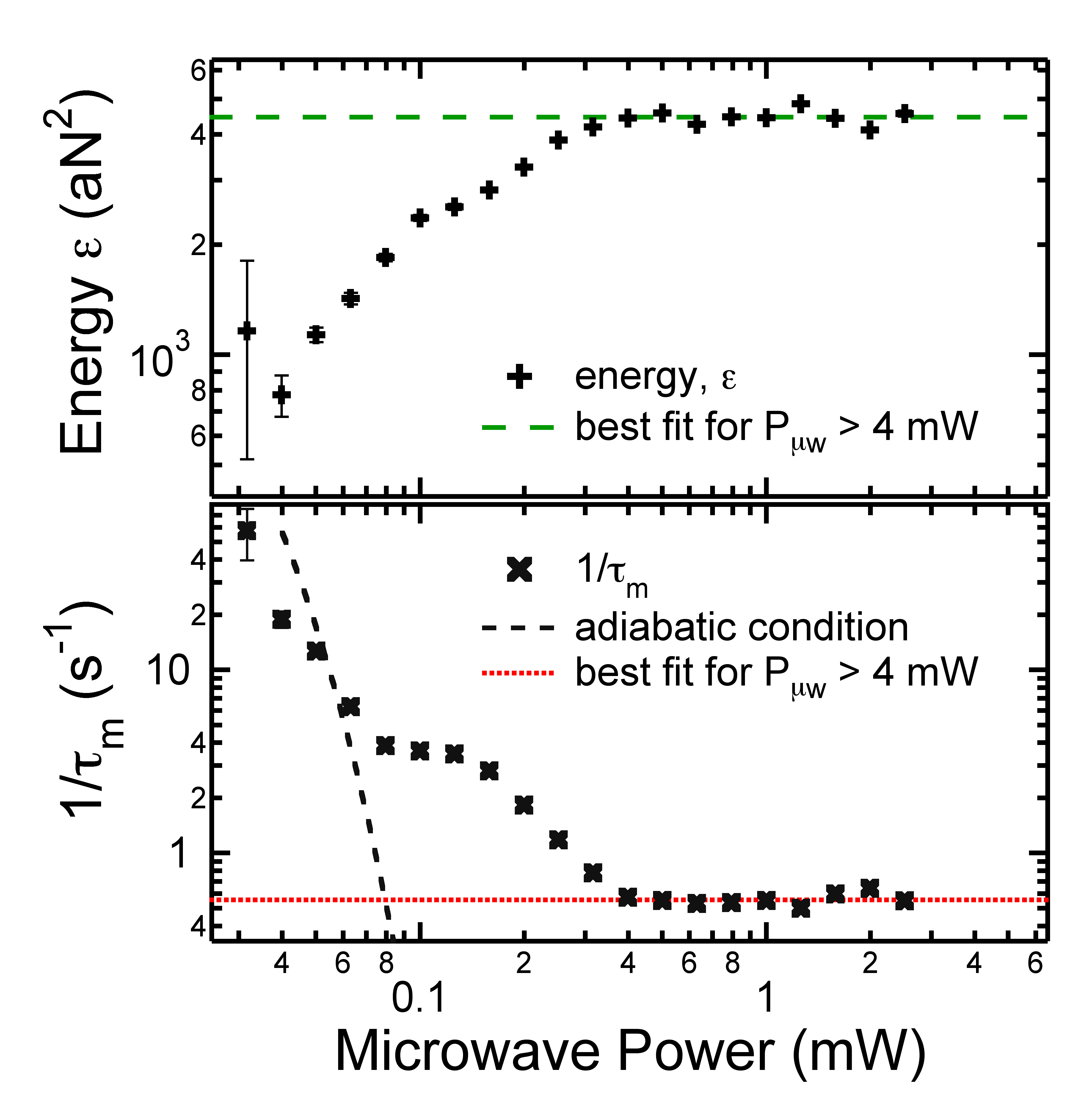}
\caption{Relaxation rate $1/\tau_m$ and signal energy $\epsilon$ versus microwave power $P_{\mu w}$. At low microwave power, extraneous mechanisms such as violation of adiabaticity (black dashed line \cite{r:CantileverInducedNuclearRelaxation, AdiabaticityParameter}) will increase $1/\tau_m$. The red dotted line shows the intrinsic $1/\tau_{m0}$  phonon mediated relaxation process. The measured signal energy $\epsilon$ saturates at an intrinsic value $\epsilon_0$ as microwave power increases (green dashed line).}
\label{fig:RelaxationNMWPower}
\end{figure}

Both violation of adiabaticity and magnetic field fluctuations due to higher order cantilever modes \cite{r:wago98, r:MaminStatisticalNMR, r:hundredspin, mozyrsky:CantRxn, r:CantileverInducedNuclearRelaxation} can limit $\tau_m$. To ensure our results are free of such artifacts, we studied the dependence of $1/\tau_m$ on microwave power $P_{\mu w}$: Fig.~\ref{fig:RelaxationNMWPower} shows $1/\tau_m$ to be independent of $P_{\mu w}$ for $P_{\mu w} > 0.4$ mW. At low power, $1/\tau_m$ increases due to violation of adiabaticity (black dashed line) or other mechanisms (the shoulder near 0.15 mW). The measured signal energy $\epsilon$ increases and saturates as $P_{\mu w}$ increases. Thus we can access a measurement parameter regime in which $\tau_m$ measures intrinsic relaxation.

\begin{figure}
\includegraphics[width=\columnwidth]{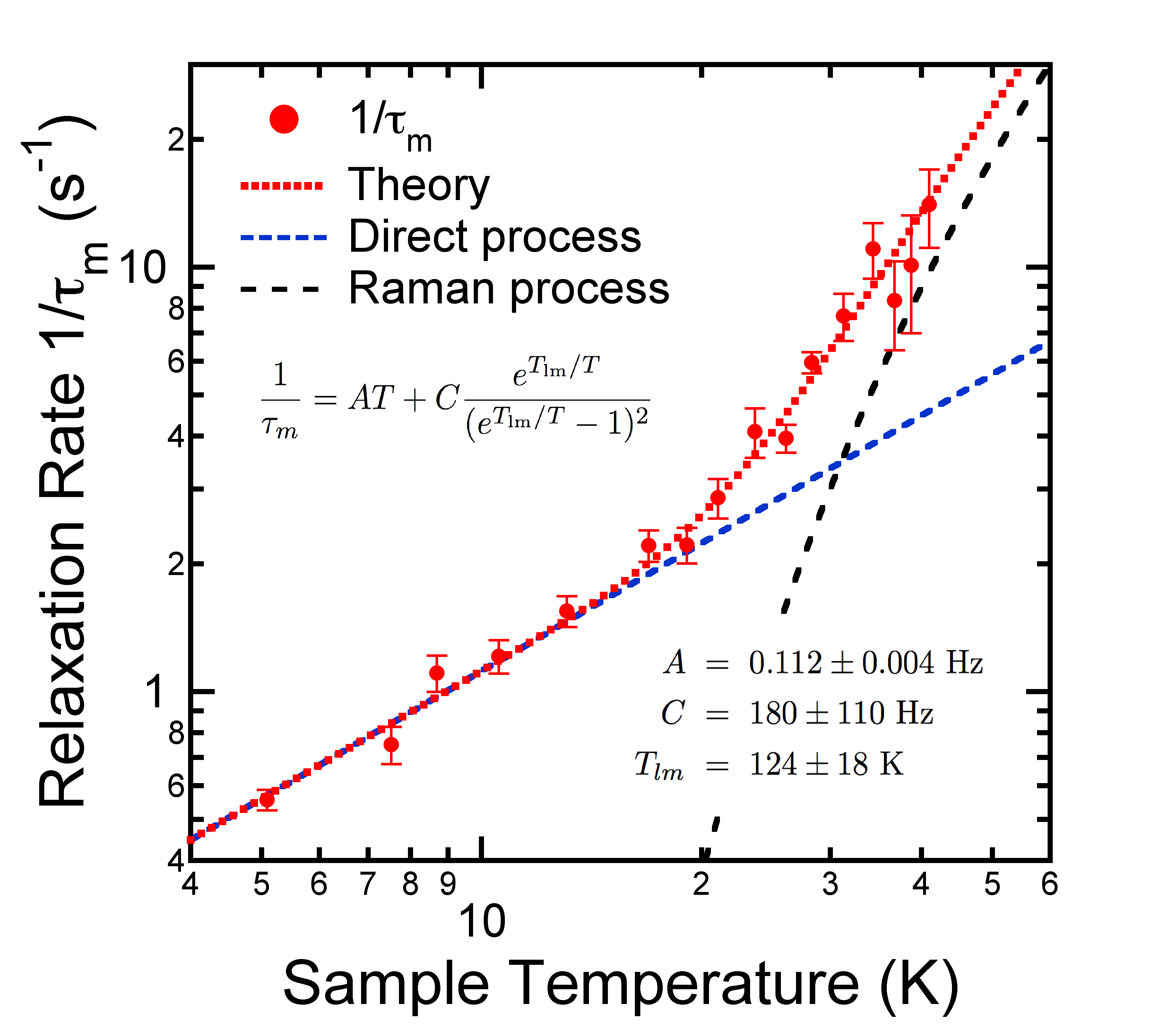}
\caption{Relaxation rate $1/\tau_m$ versus sample temperature. The red dotted line is the best fit to the direct ($AT$) and Raman local mode \( C e^ {T_{\rm lm} / T} / (e^{T_{\rm lm}/T}-1)^2 \) relaxation processes. The dashed lines show the two processes independently. Data taken with tip-sample separation $d = 200$ nm.}
\label{fig:RelaxationNTemperature}
\end{figure}
To understand the relaxation mechanism, we measured the temperature dependence of $1/\tau_m$. These measurements are presented in  Fig.~\ref{fig:RelaxationNTemperature}. To avoid spurious reduction of $\tau_m$ and thus ensure that $\tau_m(T)$ represents $T_1(T)$, $x_{\rm pk}$ and $P_{\mu w}$ were kept at 85 nm and 2.51 mW respectively. The cantilever was thermally isolated from the sample and there was no observable increase in the thermal force noise with change in sample temperature. The data are well described by the function
\begin{equation} \label{eqn:DirectLocalModeProcesses}
  AT + C \frac{e^{T_{\rm lm}/T}} {(e^{T_{\rm lm}/T} - 1)^2}
\end{equation}
where $A$, $C$ and $T_{\rm lm}$ are fitting parameters. For $T < 16$ K, direct phonon absorption or emission dominates $1/\tau_m$, and the linear temperature dependence reflects the phonon mode occupancy  $n = (\exp{(\hbar \omega_{\mu m}/k_BT)}-1)^{-1} \propto k_B T / \hbar \omega_{\mu m}$ for $T \ll \hbar \omega_{\mu m}/k_B$, where $\omega_{\mu m}$ is the microwave angular frequency, $\hbar$ and $k_B$ are the Planck and Boltzmann constants respectively. For $T >$ 16 K, the data for $1/\tau_m$ are well fit by the expression \( C e^ {T_{\rm lm} / T} / (e^{T_{\rm lm}/T}-1)^2 \) \cite{Castle:silica.jap1965, Castle:PhysRev1965}, the signature of two-phonon Raman process, in which phonons with frequency $f_{lm} = k_BT_{lm}/h$ are created and annihilated. Unlike the usual Raman process in which all available phonon modes below the Debye frequency can induce electron spin relaxation, this mechanism involves thermal excitations of the local mode of the oxygen vacancy defect only. Silica with defects induced by neutron-irradiation \cite{Shamfarov:NeutronIrradiatedQuartz, Shamfarov:37GHzQuartzData} and hydrogen impurities \cite{Castle:HydrogenImpurity, Murphy:LocalVibrationRelaxation} behave similarly. Acoustic attenuation \cite{Strakna:NeutronIrradiatedElasticModuli, Strakna:UltrasoundRelaxation}, infrared and Raman studies also support the local mode model.

Since the local mode frequencies are expected to depend on the specific type of quartz, the agreement between fitted $T_{\rm lm} = 124 \pm 18$ K and reported values \cite{Castle:EprimePhysRev1963, Castle:silica.jap1965, Castle:PhysRev1965} is satisfactory. However, the fitted value $A = 0.112 \pm 0.004$ Hz is roughly 20 times larger than that reported in Ref.~\onlinecite{Castle:silica.jap1965}. The direct process depends on the Zeeman splitting, so the discrepancy may arise from the different microwave frequencies used. We also find the scaling of $A$ deviates from the expected $\omega_{\mu w}^2$ behavior; we note similar behavior has been reported \cite{Shamfarov:37GHzQuartzData, Aminov:DirectProcessFrequencyDependence} and suggested to result from a cross relaxation process.

 \begin{figure}
 \begin{center}
 \includegraphics[width=\linewidth]{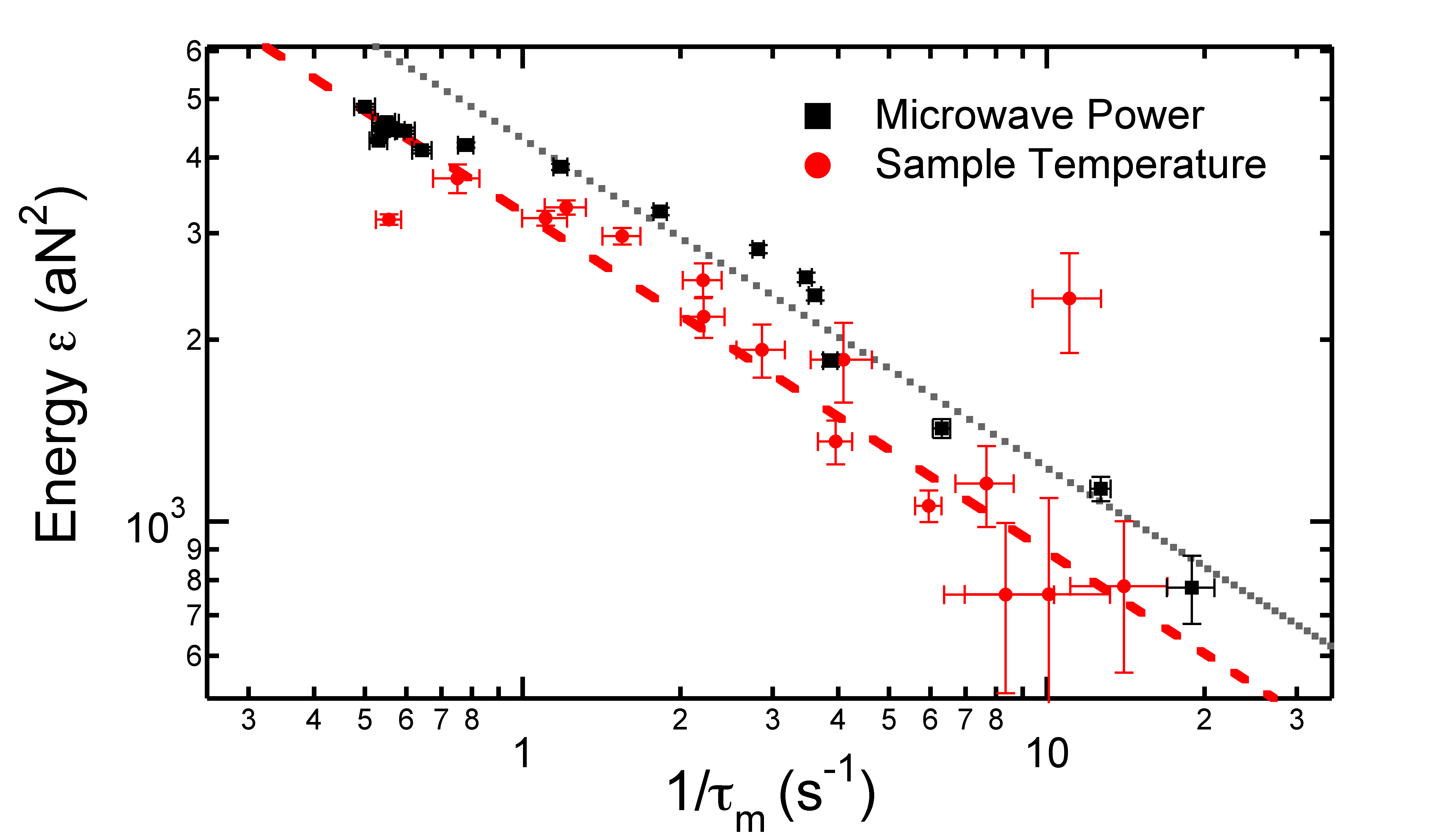}
 \caption{Power law dependence of the signal energy on $\tau_m^{-1}$: $\epsilon\propto (1/\tau_m) ^{\alpha}$ plus an offset with the sample temperature (red circles) and microwave power (black squares) as implicit parameters.  Fits to data give $\alpha = -0.56 \pm 0.05$ and $-0.54 \pm 0.04$ for varying the sample temperature and microwave power (using only $1/\tau_m > 0.7$ Hz) respectively.}
 \label{fig:EnergyvRxn}
 \end{center}
 \end{figure}

Fig.~\ref{fig:EnergyvRxn} shows a remarkable anticorrelation between $1/\tau_m$ and $\epsilon$.  When $\epsilon$ is caused to vary through its dependence on either $P_{\mu w}$ or temperature, we find \(1/\tau_m\) varies linearly with \(\epsilon ^{\alpha}\) with $\alpha \sim -0.5$. A similar dependence can be found in Ref.~\onlinecite{r:MaminStatisticalNMR}. We expect $\epsilon$ to decrease for $1/\tau_m \gg f_{m}$ due to decreased sensitivity to variations of the spin magnetization occurring within a single modulation period \cite{WindowingFunction}. This cannot explain the observed variation of $1/\tau_m$ well below $f_{m}$ (45.1 and 21.5 Hz for the $P_{\mu w}$ and temperature scans, respectively); the power $\alpha \sim -0.5$ is also not consistent with this origin. Same set of filters were used in the cantilever measurement and control circuits in these data.

We have demonstrated the ability to measure local spin relaxation times using ultrasensitive MRFM.  This points to the capability for microscopic measurement of the spatial variation of spin dynamics. This can provide new insight into materials such as superconducting cuprates where intrinsic inhomogeneity plays a central role \cite{Kivelson:StripesRMP2003}, and could provide essential understanding in technologically important phenomena such as spin coherence, spin transport and quantum information processing. Furthermore intrinsic correlation times can provide a mechanism for enhancing information content of images through relaxation rate contrast in analogy to $T_1$- and $T_2$-weighted Magnetic Resonance Imaging \cite{Haacke:MRI}.  We find, importantly, that care must be taken to understand and account for the influence of size sample size on the measured spin dynamics.

Understanding and manipulating $\tau_m$ also has implications for MRFM sensitivity. Spin noise detection SNR depends on $\tau_m$ \cite{r:NoiseNMRprl2007} because of the trade off between the averaging counts and lock-in detection bandwidth. Ref.~\onlinecite{r:NoiseNMRprl2007} uses $\pi/2$ rf pulses to randomize the spins and hence reduce $\tau_m$ to the optimal point for maximum SNR. As a consequence of the strong field gradient this required very broadband rf field which was achieved through trains of rf pulses. Our result suggest the optimal $\tau_m$ can be achieved by controlling the sample temperature.

We have studied the dependence of spin relaxation in few-electron-spin ensembles on $x_{\rm pk}$ and microwave power to understand the effect of time spent near resonance $H_1$ amplitude respectively. We have measured the intrinsic correlation time $\tau_m$ of the spin noise using the statistical polarization signal in ensembles of $\sim 100$ electron spins in vitreous silica.  Relaxation is due to coupling of spins to phonons through either a direct (single phonon) process at low temperature or a Raman process at higher temperatures. This demonstrates the capability for microscopic measurement of electron spin dynamics, an important quantity for understanding fundamental characteristics of electronic systems. Furthermore, understanding and controlling $\tau_m$ will be important for future MRFM imaging applications and sensitivity optimization.

The authors would like to thank D. Rugar and J. Mamin for providing us the sample, niobium films and lock-in software. This work was supported by the Army Research Office through MURI grant W911NF-05-1-0414.


\begin{thebibliography}{10}
\bibitem{WindowingFunction}
The lock-in detector boxcar averages the statistical polarization in a window
  $1/f_m$ long, so the suppression of signal is given by $S_{f_c}| {\rm sinc}\,
  {(\pi (f-f_m)/2f_m)}|^2$. For $1/f_m\tau \leq 1$, the attenuation is less
  than 15\%.
\end{thebibliography}
\end{document}